# Characterizing health informatics journals by subject-level dependencies: a citation network analysis

**Arezo Bodaghi**

School of Industrial and Systems Engineering,

Tarbiat Modares University, Iran

arezo.bodaghi@modares.ac.ir

**Didi Surian**

Centre for Health Informatics, Australian Institute of Health Innovation,

Macquarie University, Australia

**ABSTRACT**

Citation network analysis has become one of methods to study how scientific knowledge flows from one domain to another. Health informatics is a multidisciplinary field that includes social science, software engineering, behavioral science, medical science and others. In this study, we perform an analysis of citation statistics from health informatics journals using data set extracted from CrossRef. For each health informatics journal, we extract the number of citations from/to studies related to computer science, medicine/clinical medicine and other fields, including the number of self-citations from the health informatics journal. With a similar number of articles used in our analysis, we show that the Journal of the American Medical Informatics Association (JAMIA) has more in-citations than the Journal of Medical Internet Research (JMIR); while JMIR has a higher number of out-citations and self-citations. We also show that JMIR cites more articles from health informatics journals and medicine related journals. In addition, the Journal of Medical Systems (JMS) cites more articles from computer science journals compared with other health informatics journals included in our analysis.

# INTRODUCTION

Bibliometrics was developed to characterise and understand the inter-connectedness of large volumes of published research using statistical methods [1]. Citation analyses are a common method used in bibliometric research and cover studies that examine how authors reference prior literature, how citations correspond to the characteristics of the research, and the network structure of citation networks [2]. Health informatics is defined as a study of information and communication systems in healthcare [3]. Health informatics is a scientific discipline that handles the intersection of information science, medical informatics, computer science, and health care informatics [4].

Journals have important differences due to the existence of many research disciplines [5]. These differences are attributed to intrinsic characteristics of journal. The exchange of citations among journals forming their positions in a social structure which affect their influence[6]. Our aim was to characterize the citation structure of health informatics journals to measure differences and similarities in research focus, the coordination of research across the journals, and differences in the way the journals are informed by, and inform, medicine and computer science.

# RELATED WORK

Networks of collaboration have been investigated extensively using the network science techniques. The analysis of citation network is performed at three levels including node-level, group-level, and network-level. The node-level analysis measures the centrality of a node comprising degree, eigenvector, closeness, and betweenness [7]; the group-level analysis involves methods for detecting clusters [8]; and the network-level analysis focuses properties of networks such as distribution of node degrees [9].

A wide array of studies have considered the journal citation networks with regard to structural characteristics such as density, average and largest node distances, percolation robustness, distributions of incoming and outgoing edges, reciprocity, and assortative mixing by node degrees [10]. There are studies in which journal citation networks were analyzed empirically and focusing on communities in citation networks [11, 12]. However, most of previous studies only focused on a specific journal in the analysis.

**MATERIAL AND METHODS**

**Study data**

We selected the first ten health informatics journals ranked by Google Scholar [13] in the "medical informatics" sub-discipline. We identified 10,716 articles published in the top five health informatics journals from 1944 to 2018. From the 10,716 articles, the reference lists were available for about 1,944 articles. The information of the five health informatics journals including the digital object identifiers (DOIs) for all 1944 articles, and reference lists with the DOIs, journals' ISSN, and name for all cited references were retrieved from CrossRef ([https://www.crossref.org)](https://www.crossref.org). All journals extracted from the reference lists were labelled using CrossRef's subject list and abstracted to one of four different groups: *health informatics*, *medicine*, *computer science*, and *others*. Although some journals were listed in one subject category, the others were listed in multiple subject categories. For those journals with multiple subjects, we manually assigned them to the most relevant subject category. Currently, there are some journals that information about references and citations to CrossRef are not provided, whereas they might appear among the reference lists of articles published in journals that were included in the analysis.

**Network construction**

We generated a journal citation network from the main health informatics journals and the other extracted journals. Each journal is represented by a node and the relation between two journals is represented by an edge (a directed edge goes from an article to the article in its reference list). This network is a directed graph with 4,144 nodes (journals) and 39,656 edges among journals. Furthermore, we constructed another directed network of citations exchange among main papers for which reference lists were harvested. In this network, all 1,944 papers are considered as nodes, and edges are directed links between papers. The third network was a bipartite network comprising two types of nodes (journals and subjects) in which all five health informatics journals are on the left side, and four different subjects on the right side. In the bipartite network, there is an edge when a health informatics journal cites to a journal from a subject. Figure 1 illustrates the citation network among the journals. We used *winpython* with *networkx* and *igraph* libraries in our experiments. To construct and visualize the citation network, we used *Gephi*.

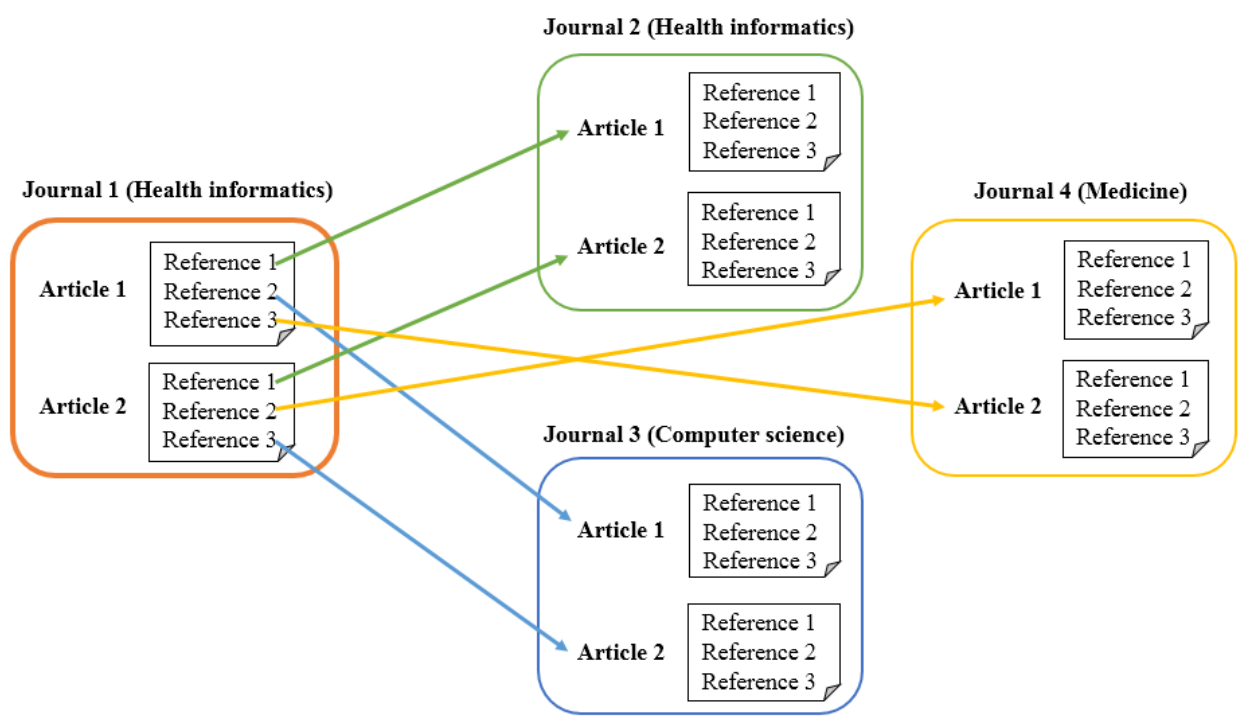

**Figure 1:** The relation among journals in four different groups: health informatics, medicine, computer science, and others.

**Analyses**

In our study, journals' overall attractiveness is measured with several measures containing incoming and outgoing citations, followed by the number of out-citations in different subjects (network and group level analysis). For investigating the role of aforementioned factors in tie-generation in the directed network of citations, various statistical terms associated with them namely in in-degree, out-degree, and loops were considered in this study. Using the number of different citations we can find out which health informatics journal receive most or least citations per paper from other journals (in-citation), and the journal that has more citations per paper to other journals (out-citations). Moreover, we can identify the health informatics journals with the highest number of citations to its papers (self-citation).

In terms of relations across subjects, we investigated the behavior of five health informatics journals (Journal of Medical Internet Research; Journal of the American Medical Informatics Association; Journal of Medical Systems; BMC Medical Informatics and Decision Making; Journal of Medical Internet Research - Mobile Health and Ubiquitous Health) in the citations to other journals of different subjects (health informatics, computer science, medicine, other fields). The number of out-citations in every different subject indicates the degree of dependence or application of different subject in health informatics.

**Results**

The section presents results of the survey. The data available for these journals varied in terms of the years for which articles were available and the years in which articles had reference list data available is shown in Table 1. In addition, the relation among main articles that the list of references are available for them, is demonstrated as a network in Figure 2.

**Table 1**: The main health informatics journals extracted from CrossRef

| Health informatics journals | Available publication years | Number of available articles | Number of articles with reference data |
|---|---|---|---|
| **Journal of Medical Internet Research (JMIR)** | 1999- 2018 | 2779 | 525 |
| **Journal of the American Medical Informatics Association (JAMIA)** | 1994- 2018 | 3021 | 524 |
| **Journal of Medical Systems (JMS)** | 1977- 2018 | 2843 | 470 |
| **BMC Medical Informatics and Decision Making (BMC MIDM)** | 2001- 2018 | 1410 | 287 |
| **Journal of Medical Internet Research - Mobile Health and Ubiquitous Health (JMU)** | 2013- 2018 | 663 | 138 |

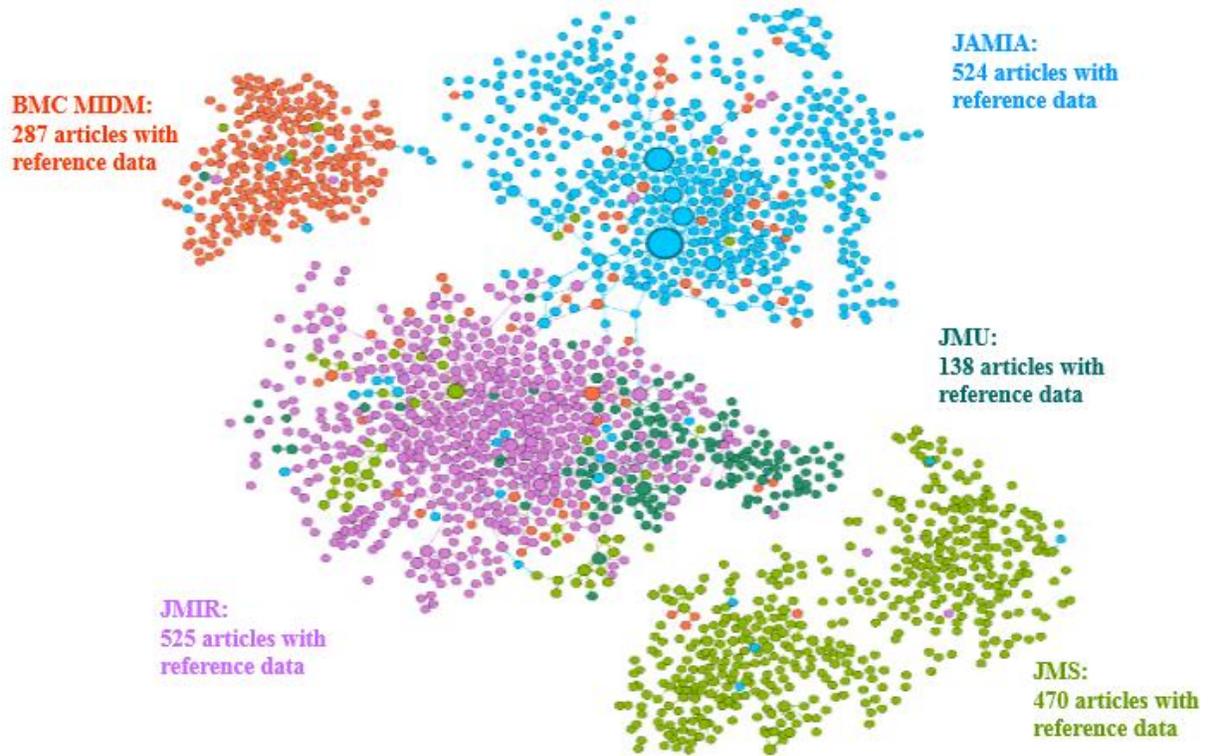

**Figure 2:** The directed network of 1,944 papers from five health informatics journals (JAMIA: blue; JMS; green; MIDM: red; JMIR: purple; JMU: dark green). Node sizes are proportional to the number of incoming citations. In this network the JAMIA cluster clearly is close to JMIR, while BMC MIDM is placed opposite to JAMIA.

The characteristics of the constructed citation network is shown in Table 2 and the citation network from the five main health informatics journals to the other journals is illustrated by Figure 3. All the main health informatics journals are positioned on the left side and the rest nodes on the right side.

**Table 2:** The characteristics of the constructed citation network

| Network properties | Results |
|---|---:|
| Number of journals (nodes) | 4,144 |
| Number of edges | 39,656 |
| Number of health informatics journals | 48 |
| Number of computer science journals | 301 |
| Number of medicine/clinical medicine journals | 912 |
| Number of other fields journals | 2883 |
| Density | 0.002 |
| Average (total) degree | 9.569 |
| Betweenness centrality | 0.00025 |
| Closeness centrality | 0.682 |
| Degree centrality | 4.219 |

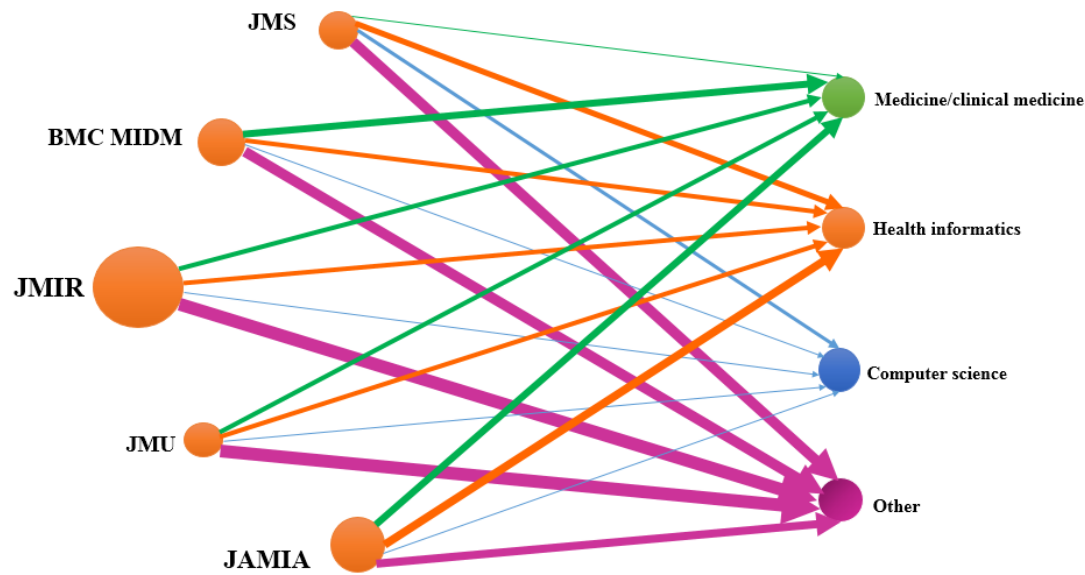

**Figure 3:** The bipartite citation network of health informatics journals extracted from CrossRef data set. The sizes of the nodes for the main health informatics journals are proportional to the number of out-citations. The sizes of the edges from one main health informatics journal are proportional to the number of the out-citations to the other journals. We merge the journals as one single node for non-main health informatics journals.

**Table 3**: The citation statistics for the main health informatics journals

|   |            | Number of papers | In-citation    | Out-citation     | Self-citation  |
|---|------------|------------------|----------------|------------------|----------------|
| 1 | **JMIR**     | 525              | 2,381 (4.53)   | 15,099 (28.76)   | 1,860 (3.54)   |
| 2 | **JAMIA**    | 524              | 2,472 (4.71)   | 8,143 (15.54)    | 1,786 (3.4)    |
| 3 | **JMS**      | 470              | 753 (1.60)     | 6,323 (13.45)    | 646 (1.37)     |
| 4 | **BMC MIDM** | 287              | 392 (1.36)     | 6,260 (21.81)    | 184 (0.64)     |
| 5 | **JMU**      | 138              | 272 (1.97)     | 3,831 (27.76)    | 167 (1.21)     |

Note: The number in the parentheses represents the average citations per paper

The distribution of out-citation and in-citation demonstrate that journals send out more citations, in comparison with receiving citations. According to Table 3, JMIR has the largest number of out-citations per paper, i.e. 28.76 (15,099 citations over 525 papers) in the network, whereas JAMIA has the largest number of in-citations per paper, i.e. 4.71 (2,472 citations over 524 papers). It shows that papers in JAMIA are more likely cited by papers from other health informatics journals. Table 3 also shows that papers in JMIR tend to cite more previously published papers in JMIR (3.54 citations in average over 525 papers).

**Table 4**: The number of out-citations from the five health informatics journals

|   |              | Out-citation     |                     |          |              |
|---|--------------|------------------|---------------------|----------|--------------|
|   |              | Computer science | Health informatics  | Medicine | Other fields |
| 1 | **JAMIA**    | 383              | 2,754               | 2,138    | 2,868        |
| 2 | **JMIR**     | 387              | 3,006               | 2,912    | 8,794        |
| 3 | **JMS**      | 806              | 1,544               | 1,052    | 2,921        |
| 4 | **BMC MIDM** | 316              | 1,225               | 1,832    | 2,887        |
| 5 | **JMU**      | 61               | 776                 | 859      | 2,135        |

Table 4 shows that papers published in JAMIA, JMIR and JMS have a likelihood to cite more papers from health informatics journals. By contrast, papers published in BMC MIDM and JMU are likely to cite more papers from medicine-related journals. Figure 4 shows a Sankey diagram which represents the out-citation flow from the main five health informatics journals (left side) to the cited journals (right side).

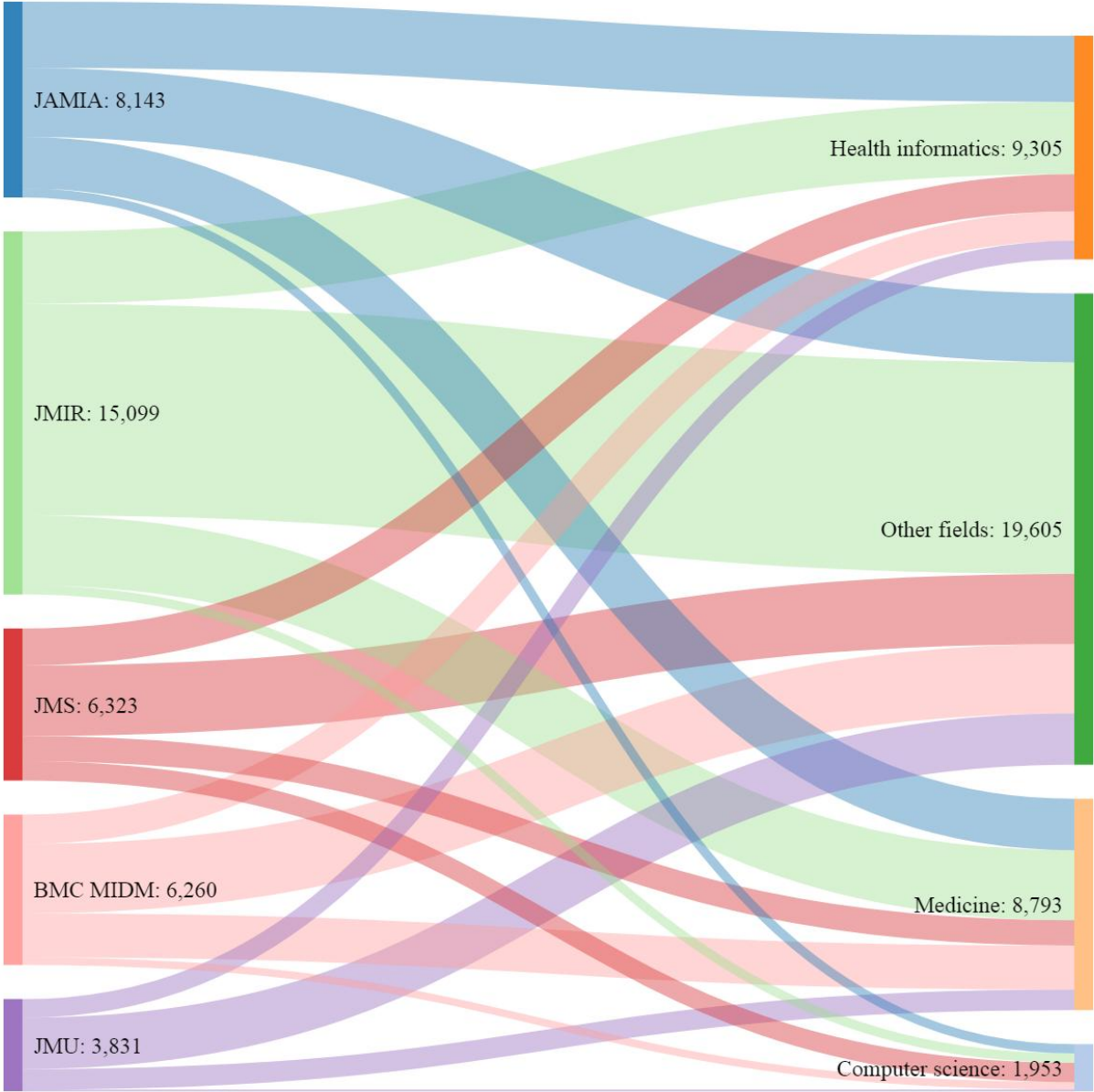

**Figure 4:** The Sankey diagram representing the citation numbers from the main health informatics journals to the cited journals.

**Discussion**

In the study, bibliometrics is employed to assess several health informatics journals in terms of number of citations. The applied mechanisms advance our understanding of the roles of references in the coordination of research across the health informatics journals and other non-health informatics journals. To the best of our knowledge, the current study is the first attempt to illustrate that the health informatics is multidisciplinary, drawing on 4 clusters of journals from health informatics, medicine, computer science and other field of studies.

There are also other methods for comparing health informatics journals which have not been applied here. One such is the examine of tie-generative mechanisms like triadic closure in forming citation links in among journals [6] which would promote our knowledge flow in this analysis a step further.

There are several limitations in this study. First, we only consider one subject (label) for each journal. The label itself is assigned by one author to each journal using semi-automate method based on the journal names and the subjects provided by the CrossRef, which may introduce a bias in the labelling and results. Another limitation is, there is inconsistency in the sampling due to incomplete data set. In this work, we rely on data set from the CrossRef and at the time of this writing, not all journals recorded on the CrossRef have complete list of articles, or complete citation information for each article in the respective journal. Finally, we also do not consider the temporal dynamics of citations of each health informatics journal. As science continually evolve, the focuses of scientific research also change from time to time [14, 15, 16], and this may contribute to the citation preferences. Rather than extracting the citation statistics from each health informatics journal on a specific time, we perform the analysis collectively from several health informatics journals published within any time range.

## Acknowledgement

We are grateful to Associate Professor Adam Dunn for suggestions and comments.